# Prediction of Silicate Glasses' Stiffness by High-Throughput Molecular Dynamics Simulations and Machine Learning

Kai Yang,[1] Xinyi Xu,[1] Benjamin Yang,[1] Brian Cook,[1] Herbert Ramos,[1] Mathieu Bauchy[1,*]

[1]Physics of AmoRphous and Inorganic Solids Laboratory (PARISlab), University of California, Los Angeles, CA 90095, U.S.A.
*Corresponding author: Prof. Mathieu Bauchy, bauchy@ucla.edu

## Abstract

The development by machine learning of models predicting materials' properties usually requires the use of a large number of consistent data for training. However, quality experimental datasets are not always available or self-consistent. Here, as an alternative route, we combine machine learning with high-throughput molecular dynamics simulations to predict the Young's modulus of silicate glasses. We demonstrate that this combined approach offers excellent predictions over the entire compositional domain. By comparing the performance of select machine learning algorithms, we discuss the nature of the balance between accuracy, simplicity, and interpretability in machine learning.

## I.   Introduction

Improving the mechanical properties of glasses is crucial to address major challenges in energy, communications, and infrastructure.[1] In particular, the stiffness of glass (e.g., its Young's modulus $E$) plays a critical role in flexible substrates and roll-to-roll processing of displays, optical fibers, architectural glazing, ultra-stiff composites, hard discs and surgery equipment, or lightweight construction materials.[1–4] Addressing these challenges requires the discovery of new glass compositions featuring tailored mechanical properties.[5,6]

Although the discovery of new materials with enhanced properties is always a difficult task, glassy materials present some unique challenges. First, a glass can be made out of virtually all the elements of the periodic table if quenched fast enough from the liquid state.[7] Second, unlike crystals, glasses are out-of-equilibrium phases and, hence, do not have to obey any fixed stoichiometry.[8] These two unique properties of glass open limitless possibilities for the development of new compositions with enhanced properties—for instance, the total number of possible glass compositions has been estimated to be around $10^{52}$![7] Clearly, only a tiny portion of the compositional envelope accessible to glass has been explored thus far.

The discovery of new glasses for a targeted application can be formulated as an optimization problem, wherein the composition needs to be optimized to minimize or maximize a cost function (e.g., the Young's modulus) while satisfying some constraints (e.g., ensuring low cost and processability).[9] Although the vast compositional envelop accessible to glass opens limitless possibilities for compositional tuning, optimization problems in such highly-dimensional spaces are notoriously challenging—which is known as the "curse of dimensionality." Namely, the virtually infinite number of possible glass compositions render largely inefficient traditional discovery methods based on trial-and-error Edisonian approaches.[10]

To overcome this challenge, accelerating the discovery requires the development of predictive models relating the composition of glasses to their engineering properties.[9] Ideally, physics-based models should offer the most robust predictions. In the case of glass stiffness, the Makishima–Mackenzie (MM) model may be the most popular model to predict the stiffness of glass.[11,12] This approach is essentially an additive model, wherein stiffness is expressed as a linear function of the oxide concentrations. However, such additive models are intrinsically unable to capture any non-linear compositional dependence, as commonly observed for stiffness.[1,5,13] On the other hand, molecular dynamics (MD) simulations offer a powerful method to compute the stiffness of a given glass.[14,15] However, MD is a brute-force method, that is, it requires (at least) one simulation per glass composition—so that the systematic use of MD to explore the large compositional envelop accessible to glass is not a realistic option.

In turn, machine learning (ML) offers an attractive and pragmatic approach to predict glasses' properties.[16] In contrast with physics-based models, ML-based models are purely data-driven and are developed by "learning" from existing databases. Although the fact that glass composition can be tuned in a continuous fashion renders glass an ideal material for ML methods, the application of ML to this material has been rather limited thus far.[16–20] This partially comes from the fact that ML methods critically relies on the existence of "useful" data. To be useful, data must be (i) available (i.e., easily accessible), (ii) complete (i.e., with a large range of parameters), (iii) consistent (i.e., obtained with the same testing protocol), (iv) accurate (i.e., to avoid "garbage in, garbage out" models), and (v) numerous (i.e., the dataset must be large). Although some glass property databases do exist,[21] some inconsistencies in the ways glasses are produced or tested among various groups may render challenging their direct use as training sets for ML methods—or would require some significant efforts in data cleaning and non-biased outlier detection.

To overcome these challenges, we present here a general method wherein high-throughput molecular dynamics simulations are coupled with machine learning methods to predict the relationship between glass composition and stiffness. Specifically, we take the example of the ternary calcium aluminosilicate (CAS) glass system—which is an archetypical model for alkali-free display glasses[22]—and focus on the prediction of their Young's modulus. We show that our method offers an excellent prediction of the Young's modulus of CAS glasses over the entire compositional domain. By comparing the performance of select ML algorithms, we show that the artificial neural network algorithm offers the highest level of accuracy. Based on these results, we discuss the balance between accuracy, complexity, and interpretability offered by each ML method.

## II. Methodology

### 1. High-throughput molecular dynamics simulations

To establish our conclusions, we first use molecular dynamics simulations to create a database consisting of the Young's modulus values of 231 glasses homogeneously covering the CAS ternary system, with 5% increments in the mol% concentration of the CaO, $Al_2O_3$, and $SiO_2$ oxide constituents. At this point, no consideration is made as to whether all these compositions would experimentally exhibit satisfactory glass-forming ability. All the simulations are conducted using the Large-scale Atomic/Molecular Massively Parallel Simulator (LAMMPS) package.[23] Each glass comprises around 3000 atoms. We adopt here the interatomic potential parametrized by Jakse—as it has been found to yield some structural and elastic properties that are in good agreement with experimental data for CAS glasses.[24,25] A cutoff of 8.0 Å is used for the short-range interactions. The Coulombic interactions are calculated by adopting the Fennell damped shifted force model with a damping parameter of 0.25 Å$^{-1}$ and a global cutoff of 8.0 Å.[26] The integration timestep is kept fixed 1.0 fs.

The glass samples are prepared with the conventional melt-quench method as described in the following.[27] First, some atoms are randomly placed in a cubic box using PACKMOL while using a distance cutoff of 2.0 Å between each atom to avoid any unrealistic overlap.[28] These initial configurations are then subjected to an energy minimization, followed by some 100 ps relaxations in the canonical (*NVT*) and isothermal-isobaric (*NPT*) ensembles at 300 K, sequentially. These samples are then fully melted at 3000 K for 100 ps in the *NVT* and, subsequently, *NPT* ensemble (at zero pressure) to ensure the loss of the memory of the initial configurations and the equilibrate the systems. Next, these liquids are cooled from 3000 K to 300 K in the *NPT* ensemble at zero pressure with a cooling rate of 1 K/ps. The obtained glass samples are further relaxed at 300 K for 100 ps in the *NPT* ensemble before the stiffness computation. Note that this quenching procedure was slightly adjusted for select compositions. First, a higher initial melting temperature of 5000 K is used for the samples wherein the $SiO_2$ concentration is larger or equal to 95%—since these glasses exhibit high glass transition temperatures. Second, a faster cooling rate of 100 K/ps is used for the samples wherein the CaO concentration is larger or equal to 90 %—as these systems would otherwise tend to crystallize with a cooling rate of 1 K/ps.

The stiffness tensor $C_{\alpha\beta}$ of the equilibrated glasses is then computed by performing a series of 6 deformations (i.e., 3 axial and 3 shear deformations along the 3 axes) and computing the curvature of the potential energy:[24,29]

$$C_{\alpha\beta} = \frac{1}{V} \frac{\partial^2 U}{\partial e_\alpha \partial e_\beta} \qquad (1)$$

where $V$ is the volume of the glass, $U$ is the potential energy, $e$ is the strain, and $\alpha$ and $\beta$ are some indexes representing each Cartesian direction. Note that all of the glass samples are found to be largely isotropic—so that the Young's modulus ($E$) can be calculated as:

$$E^{-1} = (S_{11} + S_{22} + S_{33})/3 \qquad (2)$$

where $S = C^{-1}$ is the compliance matrix.[15] Based on previous results, the Jakse forcefield is found to systematically overestimate the Young's modulus of CAS glasses by about 16%—which may be a spurious effect arising from the fast cooling rate used in MD simulations. As such, the computed Young's modulus values are rescaled by this constant factor before serving as a training set for the machine learning models presented in the following.

## 2. Machine learning methodology
### a. Inputs and outputs

The 231 Young's modulus values computed by the high-throughput MD simulations then serve as a database to infer the relationship between glass composition (*x*, *y*) and *E* in the $(CaO)_x(Al_2O_3)_y(SiO_2)_{1-x-y}$ glass system by ML. In details, we consider *x* and *y* to be the only inputs of the model (i.e., we neglect herein the effect of the thermal history of the glasses), whereas *E* is used as an output. Note that a similar approach can be used to predict the effect of composition on the shear modulus *G*, bulk modulus *K*, or Poisson's ratio *v*. In the following, we briefly describe our overall ML strategy as well as the different ML algorithms that are considered and compared herein.

### b. Data preprocessing

To avoid any risk of overfitting, a fraction of the data points is kept fully unknown to the model and is used as a "test set" to *a posteriori* assess the accuracy of each model. The test set is formed by randomly selecting 30% data points within the 231 data points of the dataset. The rest of the rest (i.e., 70%) is used as a training set, that is, to train the ML models.

Further, to obtain a proper setting for the hyperparameters of each model, a fraction of the remaining training set is kept as a "validation set." However, isolating a fixed validation set would further reduce the number of points used for training our models, which can be a serious issue in the case of a small dataset as herein. To overcome this limitation, we adopt here the *k*-fold cross-validation (CV) technique. The CV technique consists in splitting the training set into *k* smaller sets, wherein the model is trained on "*k* – 1" of the folds and validated on the remaining of the data. The results are then averaged by iteratively using each of the *k* folds for validation. Here, we use *k* = 10.

### c. Optimizing the complexity of the models

For optimal predictions, ML models must achieve the best balance between accuracy and simplicity—wherein models that are too simple are usually poorly accurate (i.e., "underfitted"), whereas models that are too complex present the risk of placing too much weight on the noise of the training set and, thereby, often show poor transferability to unknown sets of data (i.e., "overfitted"). Hence, one needs to identify the optimal degree of complexity (e.g., number of terms, number of neurons, etc.) for each model. Here, we optimize the degree of complexity of each model by gradually increasing its complexity and tracking the accuracy of the model prediction for both the training and test sets. Indeed, although the accuracy of the training set prediction typically monotonically increases with increasing model complexity, overfitted models usually manifest themselves by a decrease in the accuracy of the test set prediction (see below).

### d. Machine learning algorithms
#### i. Polynomial regression

We now detail the different learning methods used herein. We first focus on the polynomial regression (PR) method.[30] PR is a special case of multiple linear regression that includes higher degree polynomial terms and treats these higher degree polynomes as other independent variables. In general, the $N^{th}$ degree PR method can be described as:

$$Y = \beta_0 + \sum_{i=1}^{N} \beta_i X^i \qquad (3)$$

where $X$ is the input, $Y$ is the output, and the $\beta_i$ terms are the fitting parameters corresponding to each degree $i$. Here, we adopt the multivariate polynomial regression with two independent variables, which can be expressed as:

$$Y = \beta_0 + \sum_{i=1}^{N} \beta_i X_1^i + \sum_{j=1}^{N} \beta_j X_2^j + \sum_{k=1}^{N-1} \beta_k X_1^k X_2^{N-k} \qquad (4)$$

where $X_1$ and $X_2$ are the two input variables (i.e., the composition terms $x$ and $y$ herein). The least-square method is then used to identify the coefficients $\beta_i$ that minimize the sum of squared difference between the "real" values (i.e., computed by MD) and those predicted by the PR method (i.e., $Y$) during the training phase. The complexity of PR models depends on the choice of the $N^{th}$ polynomial degree considered during training.

#### ii. LASSO

One of the major disadvantages of the least-square approximation is that it tends to overfit the data used for training. LASSO (least absolute shrinkage and selection operator) regression offers a useful solution to decrease the complexity of the model and, thereby, limit the risk of overfitting.[31] This is accomplished by adding to the cost function used in PR (i.e., the sum of squared difference between "real" and "predicted" values) an additional term that further penalizes complex models. The new cost function that needs to be minimized is then defined as:

$$\sum_{i=1}^{N} \left( Y_i - \beta_0 - \sum_{j=1}^{p} X_{ij}\beta_j \right)^2 + \lambda \sum_{j=1}^{p} |\beta_j| \qquad (5)$$

where $\lambda$ is a hyperparameter that is used to control the weight attributed to the penalty associated with the complexity of the model. In practice, LASSO will force some of the $\beta_j$ parameters to be zero to minimize the value of the cost function, which results in a decrease in the complexity of the model. The degree of complexity of LASSO models can be tuned by adjusting the value of $\lambda$, namely, increasing values of $\lambda$ yield simpler models.

### iii. Random forest

The random forest (RF) method relies on the creation of a "forest," that is, an ensemble of decision trees.[32] The RF approach builds a tree by randomly choosing $n$ samples from the training set (bootstrap method). Then, at each node, it uses a randomly selected subset of variables to choose the best split to construct trees. Random forest runs input data on all $n_t$ trees and yields a prediction that is the average of all values returned by each tree:

$$Y(X) = \frac{1}{n_t} \sum_{i=1}^{n_t} Y_i(X) \tag{6}$$

where $Y_i(X)$ is the individual value predicted by one of the trees for an input vector $X$ and $Y(X)$ is the overall output of the random forest model with $n_t$ trees.[16] The degree of complexity of RF models are characterized by the number of variables in the random subset and trees in the forest.[33]

### iv. Artificial neural network

Artificial neural networks (ANN) aims to mimic the learning process of human brains. ANN models consist of an input layer that is connected to an output layer via some "hidden" layers of neurons. Each neuron takes as inputs the signals from the previous layer and produces a new output (to be used as input by the neurons from the next layer). The output $Y_i$ of a neuron $i$ in one of the hidden layers is calculated as:

$$Y_i = s(\sum_{i=1}^{N} w_i X_i + T_i^{hid}) \tag{7}$$

where $s()$ is an activation function, $N$ is the number of input neurons in the previous layer, $X_i$ are the input values, $w_i$ are the weight associated with each edge of the network, and $T_i^{hid}$ is the threshold term of hidden neurons.[16] To capture the non-linearity relationship between composition and stiffness data, we adopt herein a sigmoid function as activation function:

$$s(u) = \frac{1}{1 + e^{-u}} \tag{8}$$

We adopt here the resilient backpropagation (BP) algorithm to train the neural network model, which allows the network to learn from its errors.[34] The BP algorithm is an efficient method as it allows one to adjust the weight $w_i$ by calculating the gradient of loss function $E_{loss}$. To measure the error between the predicted and real outputs after a training sample has propagated through the network, we use the square of Euclidean distance to calculate the loss function $E_{loss}$ over $n$ training outputs as:

$$E_{loss} = \frac{1}{2n} \sum_{i=1}^{n} \left\| (Y_i - Y'_i) \right\|^2 \tag{9}$$

where $Y_i$ are the predicted outputs and $Y'_i$ is the read values (i.e., the simulated Young's modulus values). Then, after $k$ iterations, each weight $w_i$ is modified by applying an increment:

$$w_i^{(k+1)} = w_i^{(k)} + \Delta^{(k)} w_i \tag{10}$$

where $w_i^{(k+1)}$ is the updated weight, $w_i^{(k)}$ is the weight before update, and $\Delta^{(k)} w_i$ is the increment, which is calculated as by following steepest decreasing gradient in the error function as:

$$\Delta^{(k)} w_i = -\gamma_i^{(k)} sgn(\nabla_i E^{(k)}) \tag{11}$$

where $sgn()$ denotes the "sign function", $\nabla_i E^{(k)}$ denotes the partial derivative (i.e., gradient) of the error function $E^{(k)}$ with respect to weight $w_i$ at the $k^{th}$ iteration, and $\gamma_i^{(k)}$ is the learning rate at $k^{th}$ iteration. The model is iteratively refined until all the absolute values of partial derivatives of the error function become smaller than a threshold value.

e. Measuremeny of accuracy

Finally, we assess the accuracy of each model (with different degrees of complexity) by computing the RMSE (root-mean-square error) and $R^2$ factors. The RMSE factor measures the average Euclidian distance between the predicted and real values as:

$$RMSE = \sqrt{\frac{1}{n} \sum_{i=1}^{n} (Y_i - Y'_i)^2} \tag{12}$$

where $Y_i$ and $Y'_i$ are the predicted and real output values, respectively. The RMSE has the property of being in the same units as the output variables and, hence, can be used to estimate the accuracy of the Young's modulus values predicted by each model (namely, lower RMSE values indicate higher accuracy). Here, we use the RMSE of the training and test sets to determine the optimal degree of complexity for each ML model.

In complement of RMSE, we compute the $R^2$ factor, which is the percentage of the response variable variation. This factor can be used to quantify how close the data are to the fitted line. $R^2$ = 1 indicates a perfect prediction, while smaller values indicate less accurate predictions. Here, we use the $R^2$ factor to compare the performances of each different ML algorithms (once the degree of complexity has been optimized based on the RMSE).

## III. Results

### 1. Molecular dynamics simulations

We first focus on the compositional dependence of the Young's modulus values $E$ predicted by the MD simulations (see Fig. 1). Overall, we observe the existence of two main trends: (i) $E$ tends to increase with increasing $Al_2O_3$ concentration and (ii) $E$ tends to increase with increasing CaO concentration. However, we note that the dependence of $E$ on composition is non-systematic and that CaO and $Al_2O_3$ have some coupled effects. For example, we find that $E$ increases as the concentration of CaO increases when $[Al_2O_3]$ = 0 mol%, whereas $E$ decreases with increasing CaO concentration when $[Al_2O_3]$ > 40 mol%. Overall, we find that E exhibits a non-linear dependence on composition—so that one likely cannot rely on simple additive models to predict Young's modulus in the CAS system.

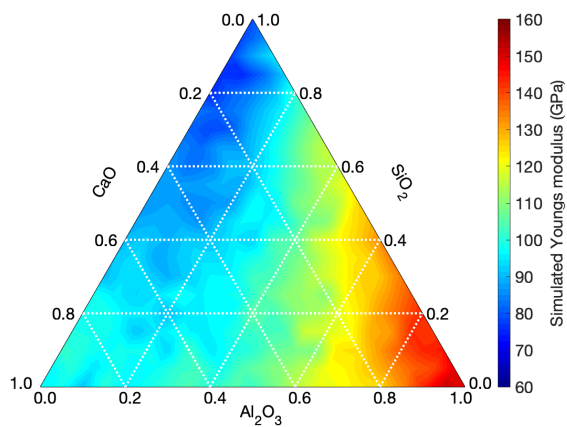

**Figure 1.** Ternary diagram showing the Young's modulus values $E$ predicted by high-throughput molecular dynamics simulations as a function of composition in the $CaO$–$Al_2O_3$–$SiO_2$ glass system. This database consists of 231 compositions homogeneously distributed over the entire compositional domain with 5 mol% increments in the oxide concentrations. This database is used as a basis to train the machine learning models presented herein.

### 2. Relationship between composition and Young's modulus

We now discuss the nature of the relationship between composition and Young's modulus. In general, the Young's modulus tends to increase with increasing connectivity.[4] To assess whether this trend is here satisfied (and whether it can be used to predict the linkage between composition and $E$), we compute based on the MD simulations the average coordination number <r> of the atoms in the network for each glass composition. As shown in Fig. 2a, we find that <r> increases with increasing CaO and $Al_2O_3$ concentrations. This arises from that fact that (i) Ca atoms have a large coordination number (around 6), while (ii) the addition of Al atoms tends to increase the degree of polymerization of the glass, i.e., by converting non-briding oxygen (NBO) into bridging oxygen (BO) atoms (we also note the formation of 5- and 6-fold overcoordinated Al species in Al-rich glasses). Overall, we obverse that the ternary plot of <r> (Fig. 2a) echoes that of $E$ (Fig. 1), which supports the fact that $E$ increases upon increasing network connectivity. Nevertheless, as shown in Fig. 2b, we find that, although $E$ and <r> are indeed positively correlated with each other, the data points are widely spread and the coefficient of determination $R^2$ only equals 0.623. This indicates that the <r> metric alone does not contain enough information to predict $E$ and that other effects are not captured by simply considering the connectivity of the network—which renders challenging the development of a robust physics-based predictive model.

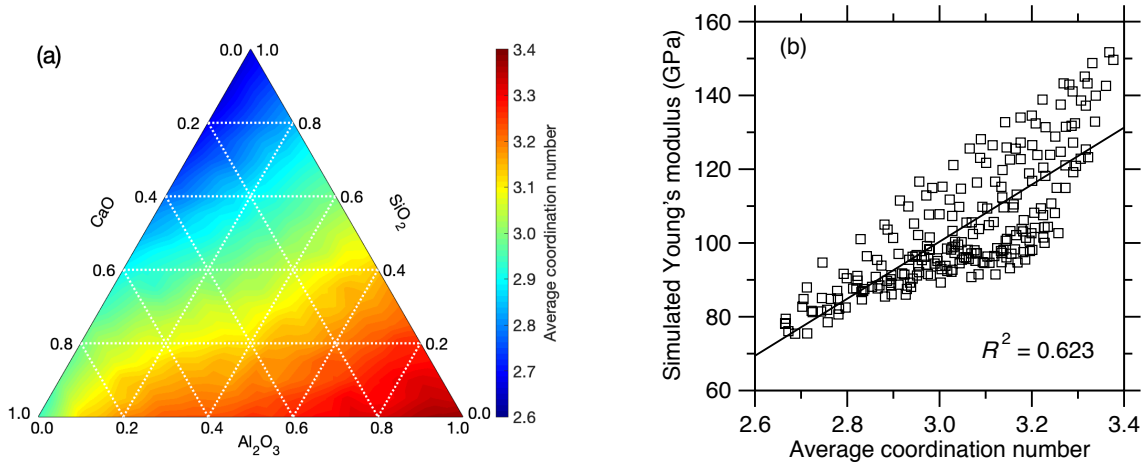

**Figure 2. (a)** Ternary diagram showing the average atomic coordination number computed by high-throughput molecular dynamics simulations as a function of composition in the $CaO$–$Al_2O_3$–$SiO_2$ glass system. **(b)** Young's modulus computed by molecular dynamics simulations as a function of the average atomic coordination number. The line is a linear fit. The coefficient of determination $R^2$ indicates the degree of linearity.

We now assess the ability of the popular Makishima–Mackenzie (MM) model to predict the compositional evolution of $E$.[11] The MM model relies on an additive relationship, wherein $E$ is expressed as a weighted average of the dissociation energies of each oxide constituent. In details, the Young's modulus $E$ is expressed as:

$$E = 83.6 \, V_t \sum_{i=1}^{n} X_i G_i \tag{13}$$

where $V_t$ is the overall packing density of the glass, and $X_i$ and $G_i$ are the concentration and volumic dissociation energy of each oxide constituent $i$, respectively. Note that the $G_i$ terms are tabulated values, whereas $V_t$ depends on the glass composition and is an explicit input of the model (i.e., the knowledge of the compositional dependence of $V_t$ is a prerequisite to the MM model). To this end, we compute the packing density $V_t$ of each glass based on the ML simulations. Figure 3a shows the ternary diagram of the $E$ values predicted by the MM model as a function of composition in the CAS glass system. We observe that the MM model properly predicts the increase of $E$ with increasing $Al_2O_3$ concentration, but fails to the increase in $E$ upon increasing CaO concentration. This is due to fact that the dissociation energy term $G$ associated with the CaO and $SiO_2$ oxides are close to each other (i.e., 15.5 and 15.4 kcal/cm³, respectively), whereas that of $Al_2O_3$ (32 kcal/cm³) is significantly higher. Overall, we observe that the MM model does not properly predict the non-linear dependence of $E$ on composition. This is not surprising as the MM is essentially an additive model (although some level of non-linearity can exist within the $V_t$ term). The MM model also fails to describe any coupling between the effects of CaO and $Al_2O_3$. Figure 3b shows a comparison between the Young's modulus values predicted by the MM model and computed by MD. Overall, we find that, although the MM model offers a fairly good prediction of $E$, the correlation remains poor (with $R^2$ = 0.556). In addition, we find that the MM model underestimates $E$, especially in the low $E$ region (which corresponds to the technologically important low-Al compositional domain wherein glasses exhibit good glass-forming ability). Overall, we note that, although the MM model can be used as a rough guide to infer some compositional trends, it cannot be used to accurately predict $E$ in CAS glasses.

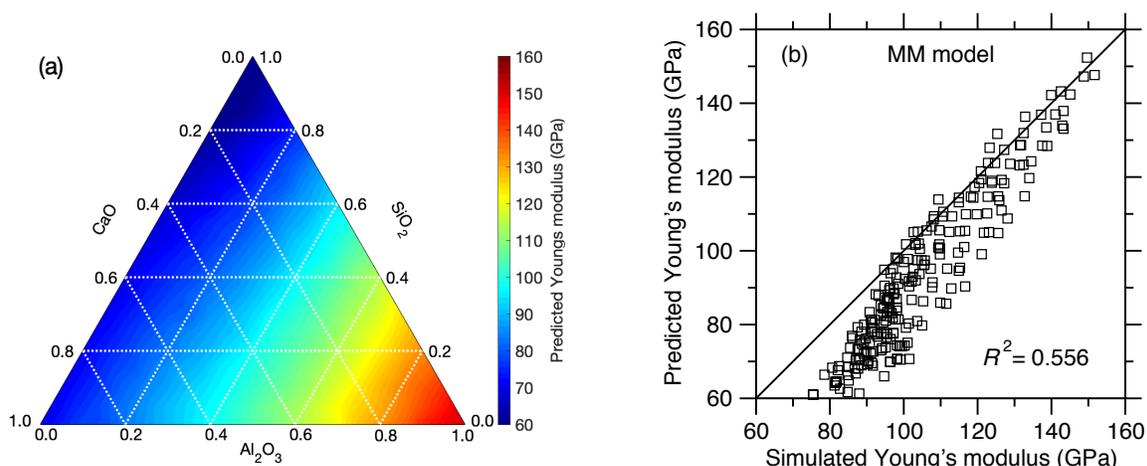

**Figure 3. (a)** Ternary diagram showing the Young's modulus values $E$ predicted by the Makishima-Mackenzie (MM) model as a function of composition in the $CaO–Al_2O_3–SiO_2$ glass system. **(b)** Comparison between the Young's modulus values predicted by the MM model and computed by molecular dynamics simulations.

### 3. Polynomial regression

Next, we assess the ability of the different ML algorithms considered herein to predict the relationship between glass composition and Young's modulus $E$. We first focus on the outcomes of polynomial regression. Figure 4a shows the RMSE offered by polynomial regression for the training and test sets as a function of the maximum polynomial degree considered in the model. As expected, we observe that the RMSE of the training set decreases upon increasing polynomial degree (i.e., increasing model complexity) and eventually plateaus. This signals that, as the model becomes more complex, it can better interpolate the training set. In contrast, we observe a significant increase in the RMSE when the polynomial degree is equal to 1 or 2—which indicates that, in this domain, the model is underfitted. This confirms again that linear models based on additive relationships are unable to properly describe the linkages between composition and Young's modulus. On the other hand, we observe that the RMSE of the test set initially decreases with increasing polynomial degree, shows a minimum for degree 3, and eventually increases with increasing degree. This demonstrates that the models incorporating some polynomial terms that are strictly larger than 3 are overfitted. This arises from the fact that, in the case of high degrees, the model starts to fit the noise of the training set rather than the "true" overall trend (see Discussion section). This exemplifies (i) how the training set allows identifying the minimum level of model complexity that is required to avoid underfitting and (ii) how the test set allows us to track the maximum level of model complexity before overfitting. Overall, the optimal polynomial degree (here found to be 3) manifests itself by a minimum in the RMSE of the test set and a plateau in the RMSE of the training set.

We now focus on assessing the accuracy of the predictions of the best polynomial regression model (i.e., with a maximum polynomial degree of 3). Figure 4b shows a comparison between the Young's modulus predicted by the ML model and computed by MD. We find that the $R^2$ factors for the training and test sets are 0.977 and 0.969, respectively. This indicates that, even in the case of a simple algorithm like polynomial regression, ML offers an excellent prediction of $E$—with an accuracy that is significantly improved with respect to that of the MM model.

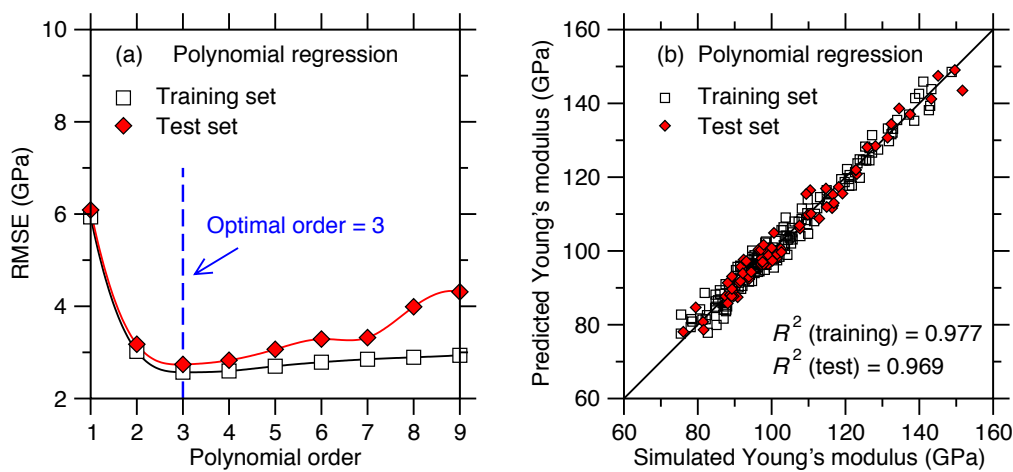

**Figure 4. (a)** Accuracy (as captured by the RMSE value) of the polynomial regression models as a function of the maximum polynomial degree considered in each model (see Sec. 2b)—as obtained for the training and test set, respectively. The optimal polynomial order is chosen as that for which the RMSE of the test set is minimum. **(b)** Comparison between the Young's modulus values predicted by polynomial regression (with a degree of 3) and computed by molecular dynamics simulations.

### 4. LASSO

We now focus on the outcomes of the LASSO algorithm. Figure 5a shows the RMSE offered by LASSO for the training and test sets as a function of the degree of complexity, $-\log(\lambda)$, of the model. In contrast with the outcomes of the polynomial regression, we observe that LASSO does not yield any noticeable overfitting at high model complexity—which would manifest itself by an increase in the RMSE of the test set. This can be understood from the fact that the LASSO algorithm specifically aims to reduce the number of polynomial terms to mitigate the risk of overfitting. Here, since the RMSE of the test set only shows a plateau with increasing $-\log(\lambda)$, we select the optimal degree of complexity as the one for which the RMSE of the test set becomes less than one standard deviation away from the minimum RMSE (i.e., in the plateau regime).

We now focus on assessing the accuracy of the predictions of the best LASSO model (i.e., with the optimal degree of complexity). Figure 5b shows a comparison between the Young's modulus predicted by the ML model and computed by MD. We find that the $R^2$ factors for the training and test sets are 0.975 and 0.970, respectively. This indicates that, here, LASSO only offers a marginal increase in accuracy as compared to polynomial regression.

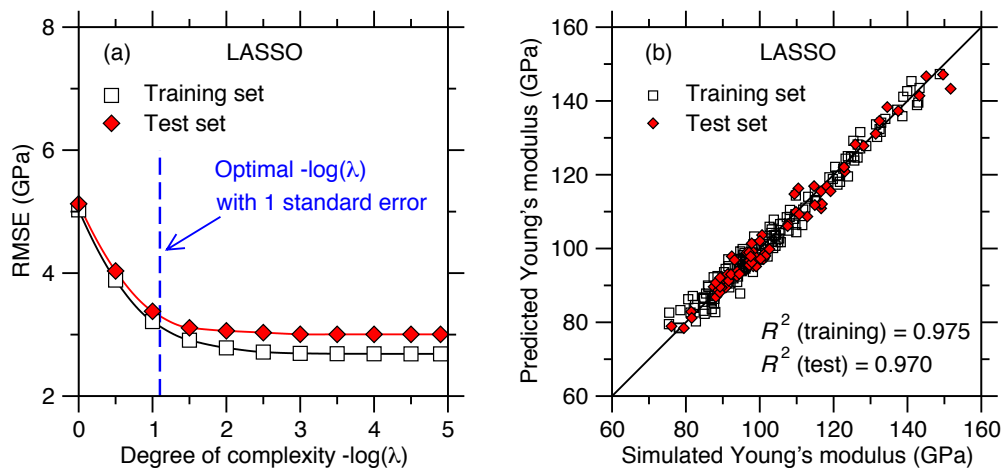

**Figure 5. (a)** Accuracy (as captured by the RMSE value) of the LASSO models as a function of the degree of complexity (see Sec. 2c)—as obtained for the training and test set, respectively. The optimal degree of complexity is determined as the one for which the RMSE of the test set is one standard deviation away from the minimum RMSE (i.e., in the plateau regime). **(b)** Comparison between the Young's modulus values predicted by LASSO (with an optimal degree of complexity) and computed by molecular dynamics simulations.

### 5. Random forest

We now focus on the outcomes of the RF algorithm. Figure 6a shows the RMSE offered by RF for the training and test sets as a function of the number of trees (i.e., which characterizes the complexity of the model). As observed in the case of LASSO, we find that RF does not yield any noticeable overfitting at high model complexity, that is, the RMSE of the test set only plateaus upon increasing number of trees. Here, we select 200 as being the optimal number of trees.

We now focus on assessing the accuracy of the predictions of the best RF model (i.e., with 200 trees). Figure 6b shows a comparison between the Young's modulus predicted by the ML model and computed by MD. We find that the $R^2$ factors for the training and test sets are 0.991 and 0.966, respectively. This suggests that, although RF offers an excellent interpolating of the training set (i.e., with a higher $R^2$ value than those obtained with the other ML models), its ability to offer a good prediction of the test set is slightly lower than those of the other ML models considered herein.

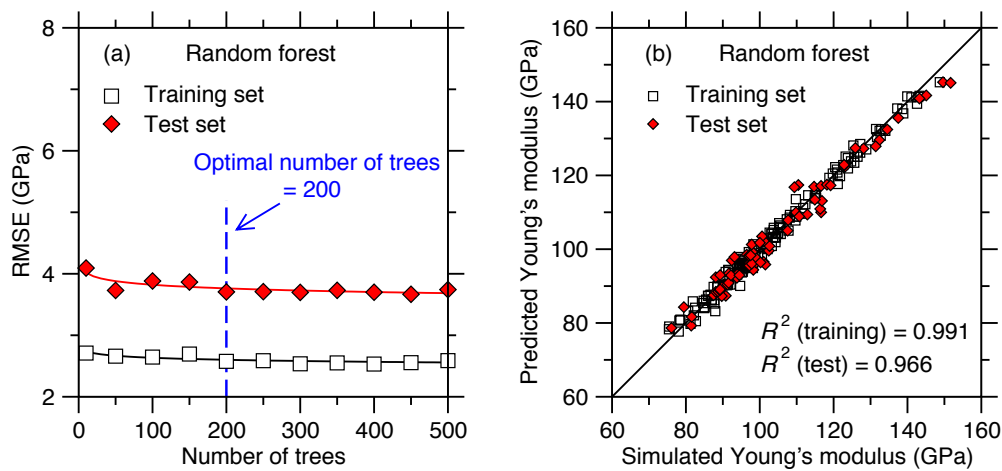

**Figure 6. (a)** Accuracy (as captured by the RMSE value) of the random forest models as a function of the number of trees considered in each model (see Sec. 2d)—as obtained for the training and test set, respectively. The optimal number of trees is taken as the threshold at which the RMSE of the test set starts to plateau. **(b)** Comparison between the Young's modulus values predicted by random forest (with 200 trees) and computed by molecular dynamics simulations.

### 6. Artificial neural network

Finally, we focus on the outcomes of the ANN algorithm. Figure 7a shows the RMSE offered by ANN for the training and test sets as a function of the number of neurons (i.e., which characterizes the complexity of the model). Overall, as previously observed in the cases of LASSO and RF, ANN does not yield any noticeable overfitting at high model complexity. Nevertheless, we note that the RMSE of the test set exhibits a slight minimum in the case of 6 neurons, which is the degree of complexity that we adopt herein.

We now focus on assessing the accuracy of the predictions of the best ANN model (i.e., with 6 neurons). Figure 7b shows a comparison between the Young's modulus predicted by the ML model and computed by MD. We find that the $R^2$ factors for the training and test sets are 0.981 and 0.974, respectively. This suggests that, although RF offers an excellent interpolating of the training set (i.e., with a higher $R^2$ value than those obtained with the other ML models), its ability to offer a good prediction of the test set is slightly lower than those of the other ML models considered herein. Overall, we find that the ANN algorithm offers the most accurate model among all the ML methods considered herein—as quantified in terms of the RMSE of the test set.

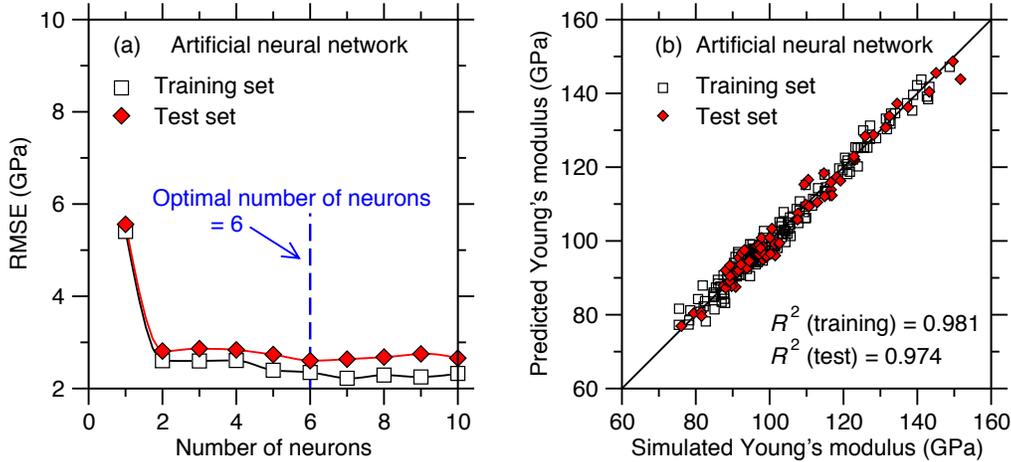

**Figure 7. (a)** Accuracy (as captured by the RMSE value) of the artificial neural network models as a function of the number of neurons considered in each model (see Sec. 2e)—as obtained for the training and test set, respectively. The optimal number of neurons is determined as that for which the RMSE value of the test set is minimum. **(b)** Comparison between the Young's modulus values predicted by artificial neural network (with 6 neurons) and computed by molecular dynamics simulations.

## IV. Discussion

### 1. Accuracy of the machine learning algorithms

We now compare the performance of the different machine learning algorithms used herein. We first focus on the level of accuracy offered by each method. To this end, Table 1 presents the coefficient of determination $R^2$ of each method for the training set (which characterizes the ability of the algorithm to properly interpolate the training data) and test set (which captures the accuracy of the model when predicting unknown data). We first observe that the RF algorithm offers the best interpolation on the training set (i.e., RF shows the highest $R^2$ for the training set). However, the RF algorithm also yields the lowest level of accuracy for the test set. This suggests that the RF algorithm presents the lowest ability to properly interpolate Young's modulus values in between two compositions of the training set and/or to offer realistic extrapolations toward the edges of the compositional domain. On the other hand, we note that the PR and LASSO algorithms offer a very similar level of accuracy, although $R^2$ offered by LASSO for the test set is slightly higher than that offered by PR. Nevertheless, we observe that the artificial neural network algorithm clearly offers the highest level of accuracy among all the models considered herein since it yields the highest $R^2$ value for the test set.

**Table 1.** Comparison between the levels of accuracy, complexity, and interpretability offered by the machine learning algorithms used herein, namely, polynomial regression (PR), LASSO, random forest (RF), artificial neural network (ANN). The level of accuracy is described by the coefficient of determination ($R^2$) for the training and test sets. The complexity is described by the number of non-zero parameters in PR and LASSO, the number of trees in RF, and the product of the number of neurons and inputs in ANN.

| ML algorithms | Coefficient of determination $R^2$ | | Complexity | Interpretability |
|:---:|:---:|:---:|:---:|:---:|
| | Training set | Test set | | |
| PR | 0.977 | 0.969 | Low (9) | High |
| LASSO | 0.975 | 0.970 | Low (8) | High |
| RF | 0.991 | 0.966 | High (200) | Intermediate |
| ANN | 0.981 | 0.974 | Intermediate (12) | Low |

To further characterize the accuracy offered by each ML algorithm, Figure 8 shows the Young's modulus values that are predicted for two series of compositions, namely, (i) $(CaO)_x(Al_2O_3)_{40-x}(SiO_2)_{60}$, wherein the $SiO_2$ fraction is kept constant and equal to 60 mol% and (ii) $(CaO)_x(Al_2O_3)_x(SiO_2)_{100-2x}$, wherein the $CaO/Al_2O_3$ molar ratio is kept constant and equal to 1. These two series specifically aim to investigate (i) the effect of the degree of polymerization of the network (i.e., fraction of NBO) and (ii) the effect of network-forming atoms (i.e., Si vs. Al) at constant degree of depolymerization (i.e., in fully-compensated glasses). We first note that, in contrast to the other ML methods, RF yields piecewise-constant-shape results, which arises from the fact that the RF method is essentially based on an ensemble of decision trees. In details, the decision tree algorithm works by relying on a binary split, that is, at each node, randomly selected observations are dropped to either the left or right daughter node depending on the values and selected features. Although a single decision tree cannot capture any non-linearity within a dataset, the output of the model is eventually averaged over all its trees—so that a RF model can capture the non-linearity of a set of data by comprising enough trees. Nevertheless, we observe here that the piecewise-constant nature of single decision trees remains encoded in the outcome of this method, which yields non-smooth predictions. This feature of the RF algorithm likely explains its excellent ability to interpolate the training set while offering only a fair prediction of the test set.

We now compare the predictions offered by the PR and LASSO algorithms. Overall, although both of the methods offer fairly comparable $R^2$ values, we note that LASSO offers an improved prediction of the $E$ at the edges of the training set (see Figs. 8a and 8b). For instance, we note that the PR method predicts an unrealistic slight increase in $E$ in pure $SiO_2$ (see the right end of Fig. 8b). This non-monotonic evolution of $E$ at the edges of the compositional domain suggests that the PR model might be slightly overfitted. In turn, such behavior is mitigated by the LASSO algorithm. Finally, we find that ANN offers the best description of the non-linear nature of the data.

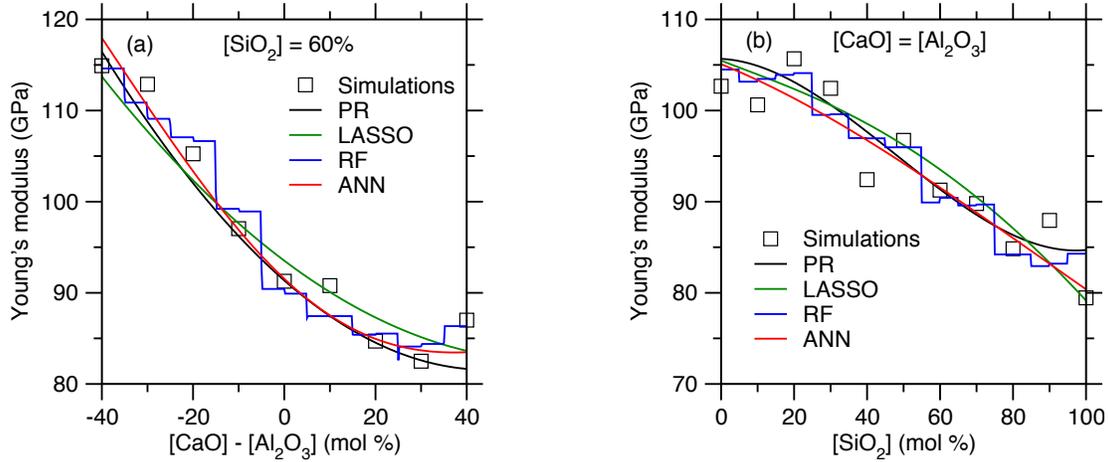

**Figure 8.** Comparison between the Young's modulus values computed by molecular dynamics simulations and predicted by the polynomial regression (PR), LASSO, random forest (RF), and artificial neural network (ANN) models for the series of compositions **(a)** $(CaO)_x(Al_2O_3)_{40-x}(SiO_2)_{60}$ and **(b)** $(CaO)_x(Al_2O_3)_x(SiO_2)_{100-2x}$.

### 2. Balance between accuracy, simplicity, and interpretability

Besides accuracy, it is also desirable for ML-based models to be "simple" (i.e., low complexity) and "interpretable" (i.e., to avoid the use of "black box" models). Unfortunately, a higher level of accuracy often comes at the expense of higher complexity and lower interpretability. Simpler and more interpretable models are usually preferable as (i) simpler models are less likely to overfit small datasets, (ii) simpler models are usually more computationally-efficient, (iii) more interpretable models are more likely to offer some new insights into the physics governing the relationship between inputs and outputs.

We now discuss the level of complexity/interpretability of the different ML-based models developed herein. The degree of complexity of each of the trained models can be roughly captured by the number of non-zero parameters in PR and LASSO, the number of trees in RF, and the product of the number of neurons and inputs in ANN (i.e., number of weight coefficients to adjust). As presented in Table 1, we first note that RF offers a poor balance between accuracy and simplicity (as the number of trees approaches the number of values in the training set). On the other hand, PR and LASSO clearly offer the lowest degree of complexity. The PR and LASSO algorithms also clearly yield the highest level of interpretability thanks to the analytical nature of the inputs/output relationship they offer. In details, we note that, although LASSO offers a level of accuracy that is fairly similar to that offered by PR, it yields a slightly simpler analytic function—with only 8 non-zero terms, vs. 9 non-zero terms for PR. This shows that, by relying on a penalized regression method, LASSO allows us to slightly enhance the level of simplicity of the model while retaining (and even slightly improving) the degree of accuracy. Finally, we note that the increased level of accuracy offered by ANN comes at the cost of higher complexity and lower interpretability, which is a common tradeoff in ML techniques.

### 3. Validation with experimental data

We now compare the predictions of the most accurate ML-based model developed herein (i.e., ANN) with the simulated data (i.e., used during the training of the model) and available experimental data (see Fig. 9).[13,35–44] We first note that the experimental data present a higher level of noise as compared to the simulation data. In the present case, these results illustrate the advantage of training ML models based on simulation rather than experimental data. Nevertheless, overall, we observe a good agreement between simulated data, ANN predictions, and experimental data. In contrast, we note that, as mentioned in Sec. III.2, the MM model systematically underestimate $E$ and does not properly capture the non-linear nature of the data. Overall, we find that the ANN model properly captures the non-linear compositional trend of $E$ while filtering out the intrinsic noise of the simulation data. These results strongly support the ability of our MD+ML combined method to offer a robust prediction of the stiffness of silicate glasses.

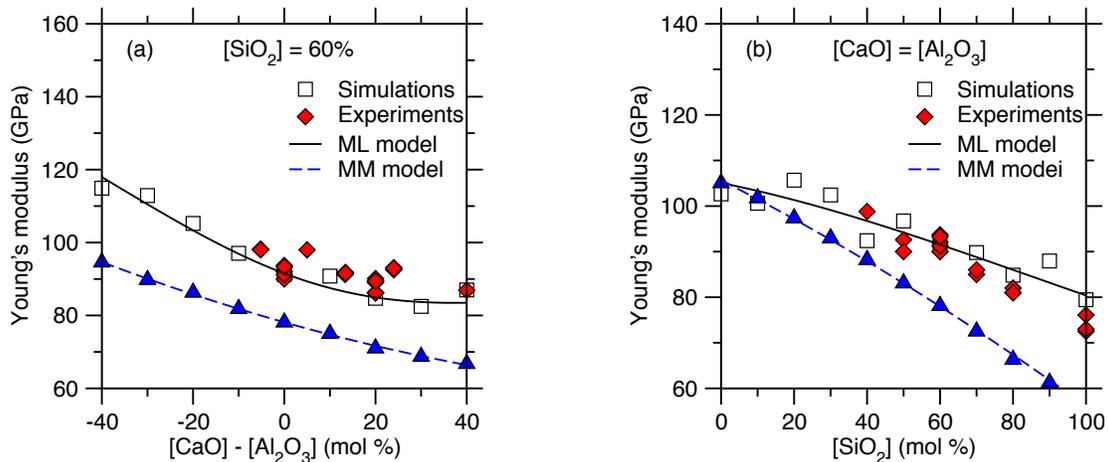

**Figure 9.** Comparison between the Young's modulus values computed by molecular dynamics simulations, predicted by the artificial neural network model, and predicted by the Makishima-Mackenzie (MM) model for the series of compositions **(a)** $(CaO)_x(Al_2O_3)_{40-x}(SiO_2)_{60}$ and **(b)** $(CaO)_x(Al_2O_3)_x(SiO_2)_{100-2x}$. The data are compared with select available experimental data.[13,35–44]

### 4. Overfitting and underfitting

We now further investigate the manifestations of underfitting and overfitting in ML-based modeling. To this end, we focus on the example of PR—since the other methods considered herein do not yield any clear signature of overfitting. To this end, Fig. 10 shows the Young's modulus values predicted by select PR models trained with varying polynomial degrees, namely, 1 (underfitted), 3 (optimal degree), and 20 (overfitted). Overall, we observe that the underfitted model (linear model with degree = 1) is obviously too simple to properly capture the non-linear nature of the dataset. In contrast, due to its high complexity, the overfitted model is able to memorize the "noise" of the simulation data used in the training set. By encoding such noise in high-degree polynomes, the overfitted model offers a poor prediction of the test set. Specifically, overfitting results in the appearance of some intense spurious ripples toward the edges of the compositional domain. Overall, the optimal model (i.e., degree 3) offers the best ability to capture the non-linearity of the data while filtering out the noise of the simulation data. These results illustrate the requirement of properly tuning the level of complexity of ML models, as illustrated by our approach.

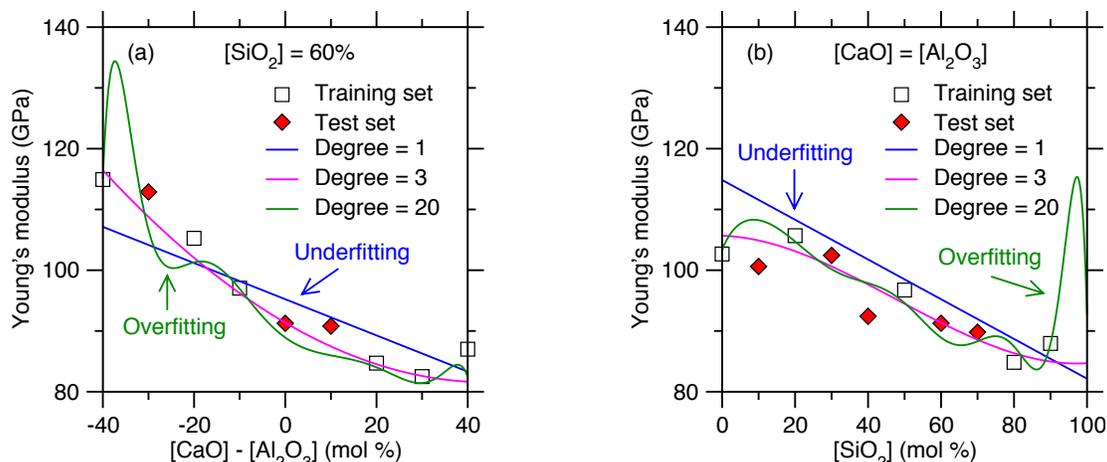

**Figure 10.** Comparison between the Young's modulus values computed by molecular dynamics simulations (wherein the training and test sets are indicated as white and red symbols, respectively) and predicted by select polynomial regression models with polynomial degrees of 1 (underfitted), 3 (optimal), and 20 (overfitted) for the series of compositions **(a)** $(CaO)_x(Al_2O_3)_{40-x}(SiO_2)_{60}$ and **(b)** $(CaO)_x(Al_2O_3)_x(SiO_2)_{100-2x}$.

## 4. Advantages of combining physics-driven (MD) and data-driven (ML) modeling

Finally, we discuss the advantages of combining ML with high-through MD simulations—rather than directly training ML-based on available experimental data. First, we note that, although the CAS ternary system may be one of the most studied systems in glass science and engineering, the number of available experimental stiffness data available for this system is fairly limited. Further, most of the data available for this system are clustered in some small regions of the whole compositional domain (namely, pure silica, per-alkaline aluminosilicates, and calcium aluminates glasses). Such clustering of the data is a serious issue as, in turn, available experimental data come with a notable uncertainty—for instance, the Young's modulus of select glasses (at fixed composition) can vary by as much as 20 GPa among different references. As such, the combination of a high level of noise and clustering of the data would not allow ML to discriminate the "true" trend of the data from the noise. Finally, we note that conducting MD simulations is obviously faster/cheaper than synthesizing glass samples and measuring their stiffness. In turn, the results presented herein demonstrate that properly conducted MD simulations can offer a quantitative agreement with experimental data and, thereby, offer a desirable alternative to systematic experiments.

# V. Conclusions

Overall, these results demonstrate that the combination of high-throughput molecular dynamics simulations and machine learning offers a robust approach to predict the mechanical properties of silicate glasses. Further, our method clearly identifies the optimal level of complexity of each ML-based model, that is, to mitigate the risk of under- or overfitting. Based on these results, we find that the artificial neural network algorithm offers the highest level of accuracy. In contrast, the LASSO algorithm offers a model that features higher simplicity and interpretability—at the expense of a slight decrease in accuracy. The method presented herein is generic and transferable to new properties (e.g., other stiffness metrics) and new systems (e.g., other families of silicate glasses).

# Acknowledgments

This work was supported by the National Science Foundation under Grants No. 1562066, 1762292, and 1826420.

# References

[1] L. Wondraczek, J.C. Mauro, J. Eckert, U. Kühn, J. Horbach, J. Deubener, and T. Rouxel, Adv. Mater. **23**, 4578 (2011).
[2] T. Rouxel, in *Challenging Glass Conf. Archit. Struct. Appl. Glass Fac. Archit. Delft Univ. Technol. May 2008* (IOS Press, 2008), p. 39.
[3] T. Rouxel, Comptes Rendus Mécanique **334**, 743 (2006).
[4] T. Rouxel, J. Am. Ceram. Soc. **90**, 3019 (2007).
[5] J.C. Mauro, C.S. Philip, D.J. Vaughn, and M.S. Pambianchi, Int. J. Appl. Glass Sci. **5**, 2 (2014).
[6] J.C. Mauro and E.D. Zanotto, Int. J. Appl. Glass Sci. **5**, 313 (2014).
[7] E.D. Zanotto and F.A.B. Coutinho, J. Non-Cryst. Solids **347**, 285 (2004).
[8] A.K. Varshneya, *Fundamentals of Inorganic Glasses* (Academic Press Inc, 1993).
[9] J.C. Mauro, Curr. Opin. Solid State Mater. Sci. **22**, 58 (2018).
[10] H. Liu, T. Du, N.M.A. Krishnan, H. Li, and M. Bauchy, Cem. Concr. Compos. (2018).
[11] A. Makishima and J.D. Mackenzie, J. Non-Cryst. Solids **12**, 35 (1973).
[12] A. Makishima and J.D. Mackenzie, J. Non-Cryst. Solids **17**, 147 (1975).
[13] R.J. Eagan and J.C. Swearekgen, J. Am. Ceram. Soc. **61**, 27 (1978).
[14] J. Du, in *Mol. Dyn. Simul. Disord. Mater. Netw. Glas. Phase-Change Mem. Alloys*, edited by C. Massobrio, J. Du, M. Bernasconi, and P.S. Salmon (Springer International Publishing, Cham, 2015), pp. 157–180.
[15] A. Pedone, G. Malavasi, A.N. Cormack, U. Segre, and M.C. Menziani, Chem. Mater. **19**, 3144 (2007).
[16] N.M. Anoop Krishnan, S. Mangalathu, M.M. Smedskjaer, A. Tandia, H. Burton, and M. Bauchy, J. Non-Cryst. Solids **487**, 37 (2018).
[17] C. Dreyfus and G. Dreyfus, J. Non-Cryst. Solids **318**, 63 (2003).

[18] J.C. Mauro, A. Tandia, K.D. Vargheese, Y.Z. Mauro, and M.M. Smedskjaer, Chem. Mater. **28**, 4267 (2016).
[19] M.C. Onbaşlı, A. Tandia, and J.C. Mauro, Handb. Mater. Model. 1 (2018).
[20] D.R. Cassar, A.C.P.L.F. de Carvalho, and E.D. Zanotto, Acta Mater. **159**, 249 (2018).
[21] A.I. Priven and O.V. Mazurin, Adv. Mater. Res. **39–40**, 145 (2008).
[22] A. Ellison and I.A. Cornejo, Int. J. Appl. Glass Sci. **1**, 87 (2010).
[23] S. Plimpton, J. Comput. Phys. **117**, 1 (1995).
[24] M. Bauchy, J. Chem. Phys. **141**, 024507 (2014).
[25] M. Bouhadja, N. Jakse, and A. Pasturel, J. Chem. Phys. **138**, 224510 (2013).
[26] C.J. Fennell and J.D. Gezelter, J. Chem. Phys. **124**, 234104 (2006).
[27] X. Li, W. Song, K. Yang, N.M.A. Krishnan, B. Wang, M.M. Smedskjaer, J.C. Mauro, G. Sant, M. Balonis, and M. Bauchy, J. Chem. Phys. **147**, 074501 (2017).
[28] L. Martínez, R. Andrade, E.G. Birgin, and J.M. Martínez, J. Comput. Chem. **30**, 2157 (2009).
[29] H. Liu, S. Dong, L. Tang, N.M.A. Krishnan, G. Sant, and M. Bauchy, J. Mech. Phys. Solids **122**, 555 (2019).
[30] S.M. Stigler, J. Am. Stat. Assoc. **66**, 311 (1971).
[31] R. Tibshirani, J. R. Stat. Soc. Ser. B Methodol. **58**, 267 (1996).
[32] L. Breiman, Mach. Learn. **45**, 5 (2001).
[33] A. Liaw and M. Wiener, Forest **23**, (2001).
[34] D.E. Rumelhart, G.E. Hinton, and R.J. Williams, Nature **323**, 533 (1986).
[35] C. Ecolivet and P. Verdier, Mater. Res. Bull. **19**, 227 (1984).
[36] S. Inaba, S. Todaka, Y. Ohta, and K. Morinaga, J. Jpn. Inst. Met. **64**, 177 (2000).
[37] S. Inaba, S. Oda, and K. Morinaga, J. Jpn. Inst. Met. **65**, 680 (2001).
[38] C. Weigel, C. Le Losq, R. Vialla, C. Dupas, S. Clément, D.R. Neuville, and B. Rufflé, J. Non-Cryst. Solids **447**, 267 (2016).
[39] J. Rocherulle, C. Ecolivet, M. Poulain, P. Verdier, and Y. Laurent, J. Non-Cryst. Solids **108**, 187 (1989).
[40] M. Yamane and M. Okuyama, J. Non-Cryst. Solids **52**, 217 (1982).
[41] S. Sugimura, S. Inaba, H. Abe, and K. Morinaga, J. Ceram. Soc. Jpn. **110**, 1103 (2002).
[42] T.M. Gross, M. Tomozawa, and A. Koike, J. Non-Cryst. Solids **355**, 563 (2009).
[43] *Handbook of Glass Properties* (Elsevier, 1986).
[44] I. Yasui and F. Utsuno, in *Comput. Aided Innov. New Mater. II*, edited by M. Doyama, J. Kihara, M. Tanaka, and R. Yamamoto (Elsevier, Oxford, 1993), pp. 1539–1544.